# Spin filtering in CrI$_3$ tunnel junctions

Tula R. Paudel and Evgeny Y. Tsymbal

*Department of Physics and Astronomy & Nebraska Center for Materials and Nanoscience*
*University of Nebraska, Lincoln, Nebraska 68588, USA*

The recently discovered magnetism of two-dimensional (2D) van der Waals crystals have attracted a lot of attention. Among these materials is CrI$_3$ – a magnetic semiconductor exhibiting transitions between antiferromagnetic and ferromagnetic orderings under the influence of an applied magnetic field. Here, using first-principles methods based on density functional theory, we explore spin-dependent transport in tunnel junctions formed of fcc Cu (111) electrodes and a CrI$_3$ tunnel barrier. We find about 100% spin polarization of the tunneling current for a ferromagnetically-ordered four-monolayer CrI$_3$ and tunneling magnetoresistance of about 3,000% associated with a change of magnetic ordering in CrI$_3$. This behavior is understood in terms of the spin and wave-vector dependent evanescent states in CrI$_3$ which control the tunneling conductance. We find a sizable charge transfer from Cu to CrI$_3$ which adds new features to the mechanism of spin-filtering in CrI$_3$-based tunnel junctions. Our results elucidate the mechanisms of spin filtering in CrI$_3$ tunnel junctions and provide important insights for the design of magnetoresistive devices based on 2D magnetic crystals.

**Introduction**

The growing family of two-dimensional (2D) materials offers a platform to design novel van der Waals heterostructures with unique electronic and optical properties.[1,2] Recently, the range of these properties has been expanded to include 2D magnetism.[3,4,5] Although a magnetic order is prohibited within the 2D isotropic Heisenberg model at finite temperatures,[6] magnetocrystalline anisotropy lifts this restriction and enables the appearance of the 2D Ising ferromagnetism. The discovery of long-range magnetic order in 2D crystals, such as CrI$_3$,[3] opens a new route for the fundamental studies of spin behavior at low dimensions and could lead to novel spintronics applications.

CrI$_3$ is a semiconductor with easily exfoliating monolayers. The van der Waal's force binds stiff CrI$_3$ monolayers with a cleavage energy comparable to that of transition metal chalcogenides and graphene.[7,8] Bulk CrI$_3$ is a ferromagnet with the Curie temperature of 61 K, the magnetic moment of 3.1$\mu_B$ per Cr atom[9] being oriented out of the layer plane with magnetic anisotropy of about 650 µeV.[10] When exfoliated, CrI$_3$ exhibits a magnetic order dependent on layer thickness: whereas a CrI$_3$ monolayer is an Ising ferromagnet, a CrI$_3$ bilayer is an antiferromagnet consisting of two ferromagnetic monolayers with antiparallel magnetic moments.[3] Under an applied magnetic field of about 1T, a CrI$_3$ bilayer exhibits a spin-flip transition, which changes the interlayer magnetic alignment from antiferromagnetic (AFM) to ferromagnetic (FM). This indicates a relatively weak interlayer exchange coupling of about 0.2 meV, which allows tuning the spin-flip transition by electric means.[11] The switching of the magnetic order in CrI$_3$ from AFM to FM has recently been demonstrated by the electric field effect [12] and electrostatic doping[11,13] under a constant magnetic bias.

The existence of magnetism in CrI$_3$, which can be engineered in a layer-by-layer fashion, expands functional properties of the 2D materials. It makes possible to design van der Waal's heterostructures combining magnetic and non-magnetic components where the electrical and optical properties can be controlled by magnetic ordering of the adjacent layers. A prominent example is a large tunneling magnetoresistance (TMR) effect of 95%, 300%, and 550% which was reported for graphite/CrI$_3$/graphite tunnel junctions with bilayer, trilayer, and tetralayer CrI$_3$ barriers.[14] Here, the CrI$_3$ served as a tunneling barrier and TMR was associated with a change of magnetic ordering of CrI$_3$ from AFM to FM under the influence of an applied magnetic field. This is different from the conventional TMR effect in magnetic tunnel junctions where two FM electrodes are realigned by an applied magnetic field, resulting in a change of tunneling resistance.[15] Even higher values of TMR were demonstrated for other types of tunnel junctions utilizing CrI$_3$ barrier layers of different thickness: 19,000%,[16] 10,000%,[17] and even 1,000,000%.[18]

The appearance of the giant TMR effect in tunnel junctions based on CrI$_3$ barriers is understood in terms of the spin-filtering effect.[14-18] According to the density-functional theory (DFT) calculations, the conduction band of CrI$_3$ is fully spin polarized,[7,19] so that the band gap for majority-spin electrons is smaller than that for minority-spin electrons. This creates a spin-dependent tunneling barrier which filters electrons according to their spin orientation. Therefore, when the CrI$_3$ monolayers are FM-ordered under the influence of a sufficiently large magnetic field, the tunneling current is dominated by majority-spin electrons which encounter a low tunneling barrier, and hence the conductance is high. However, when the CrI$_3$ monolayers are AFM-ordered at zero magnetic field, both majority- and minority-spin electrons encounter a higher barrier, which height varies across the CrI$_3$ layer, and the conductance is low. Note, that the detection of the spin-filtering effect in CrI$_3$-based tunnel junctions is different from that in the conventional spin-filter tunnel junctions, where a FM counter electrode is required to select the tunneling spins.[20,21]

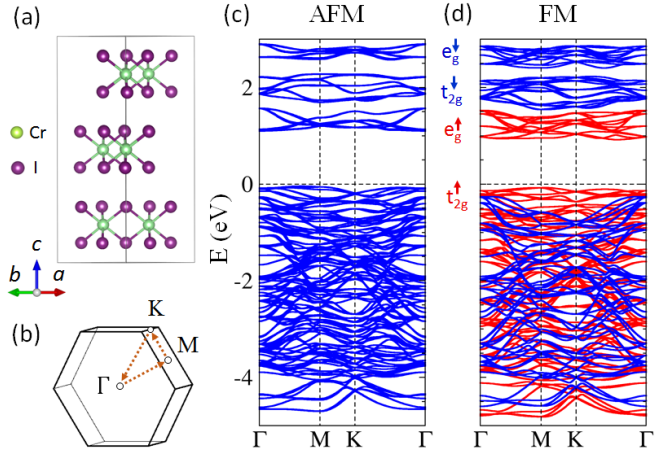

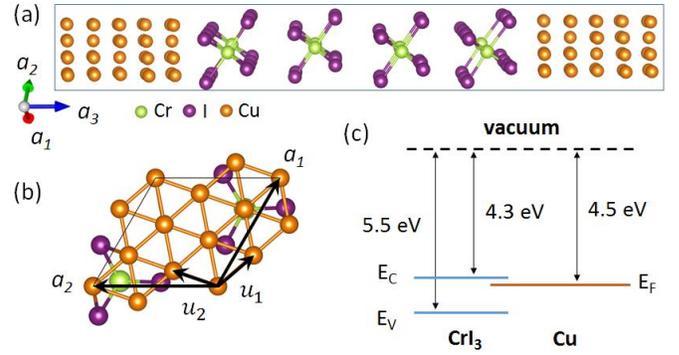

**Fig. 1** Atomic and electronic structure of bulk $CrI_3$. Atomic structure of the unit cell of bulk $CrI_3$ (a) and the corresponding Brillouin zone with the high-symmetry points indicated (b). Spin projected band structure of bulk AFM (c) and FM (d) $CrI_3$ calculated along the high symmetry lines indicated in (b) with the Cr orbital character for selected bands shown. Red (blue) color in (d) designates the majority (minority)-spin bands.

**Fig. 2** Side (a) and top (b) views of the atomic structure of the Cu(5ML)/$CrI_3$(4ML)/Cu(5ML) tunnel junction. (c) Band alignment between bulk $CrI_3$ (001) and fcc Cu (111).

Although qualitatively the mechanism of the giant TMR in $CrI_3$-based tunnel junctions has been articulated, the quantitative analysis and detailed physical understanding of the spin-filtering effect is still missing. In this paper, we employ DFT calculations to elucidate the spin-dependent transport properties of the tunnel junctions, which consist of fcc Cu (111) electrodes and a magnetic $CrI_3$ barrier layer. We find a nearly 100% spin polarization of the tunneling current for ferromagnetically-ordered $CrI_3$ and a giant TMR effect increasing with $CrI_3$ thickness. This behavior is explained in terms of the spin and wave-vector dependent evanescent states in $CrI_3$ which control the tunneling conductance. In addition, we find a sizable charge transfer from Cu to $CrI_3$ which adds new features to the mechanism of spin-filtering in $CrI_3$-based tunnel junctions.

### Results
#### Bulk $CrI_3$

Bulk $CrI_3$ in hexagonal setting of space group 148 ($R\bar{3}$) contains three $CrI_3$ layers stacked in the [0001] direction as shown in Figure 1a. Figures 1c-d show the calculated band structure of bulk $CrI_3$ along the high-symmetry directions in the 3D Brillion zone (Fig. 1b) for AFM- (Fig. 1c) and FM- (Fig. 1d) ordered $CrI_3$. The calculated energy gap is 0.92 eV for FM and 1.05 eV for AFM configuration. Both values are smaller than the measured band gap of 1.2 eV,[9] which is due to well-known deficiency of the standard DFT approach.[22] The energy position and the orbital character of the conduction band minimum (CBM) control transport properties of $CrI_3$. The band structure of $CrI_3$ is determined by the crystal field splitting of the 3d orbitals of the $Cr^{3+}$ ion ([Ar] $3d^3$) located at the center of the octahedron formed by the six $I^{-1}$ ions (Fig. 1a). The crystal field of octahedral symmetry breaks the Cr 3d orbital state into the triply degenerate $t_{2g}$ state, which lies at a lower energy, and the doubly degenerate $e_g$ state, which lies at a higher energy. According to the Hund's rule, the majority-spin $t_{2g}$ states are fully occupied by the three available $Cr^{+3}$ electrons and form the valence band, while the minority-spin $t_{2g}$ states and the $e_g$ states are empty and form the conduction band (Fig. 1d). The exchange splitting of the spin bands makes the band gap in $CrI_3$ spin-dependent. While the majority-spin CBM is formed of the $e_g$ states and lies at about 1 eV above the valence band maximum (VBM), the minority-spin CBM is formed of the $t_{2g}$ states and lies at about 1.6 eV above the VBM (Fig. 1d). The spin-dependent band gap is responsible for the spin-dependent barrier height in tunnel junctions based on $CrI_3$, which is the origin of the spin-filtering effect.

#### Cu/$CrI_3$/Cu tunnel junction: Atomic structure

We consider a Cu/$CrI_3$/Cu tunnel junction which consists of a 4-monolayer $CrI_3$ (0001) tunnel barrier placed between the 5 monolayers of Cu (111) as shown in Fig 2a. We assume the in-plane constant of the supercell to be the theoretical bulk lattice constant of $CrI_3$, $a = 6.81$ Å. In the supercell, the lattice vectors corresponding to the $CrI_3$ slab are $\vec{a}_1 = (a/2, \sqrt{3}a/2, 0)$, $\vec{a}_2 = (a/2, -\sqrt{3}a/2, 0)$ and $\vec{a}_3 = (0, 0, c)$, which represent the unit cell of bulk hexagonal $CrI_3$. To match the fcc Cu electrodes to the $CrI_3$ slab, we enlarge the in-plane Cu lattice parameters to be $\vec{a}_1 = 3\vec{u}_1 + \vec{u}_2$ and $\vec{a}_2 = 2\vec{u}_2 - \vec{u}_1$, where $\vec{u}_1$ and $\vec{u}_2$ are the unit vectors in the Cu (111) plane (Fig. 2b). The estimated in-plane strain between the Cu and $CrI_3$ layers is $\epsilon_{11} = \epsilon_{22} \approx 1\%$ [23] suggesting a possibility of good epitaxy. The Cu layer is positioned in such a way that one of the Cu atoms lies directly above the Cr atom, as shown in Fig. 2b, so that it has the same distance from the nearest I atoms.

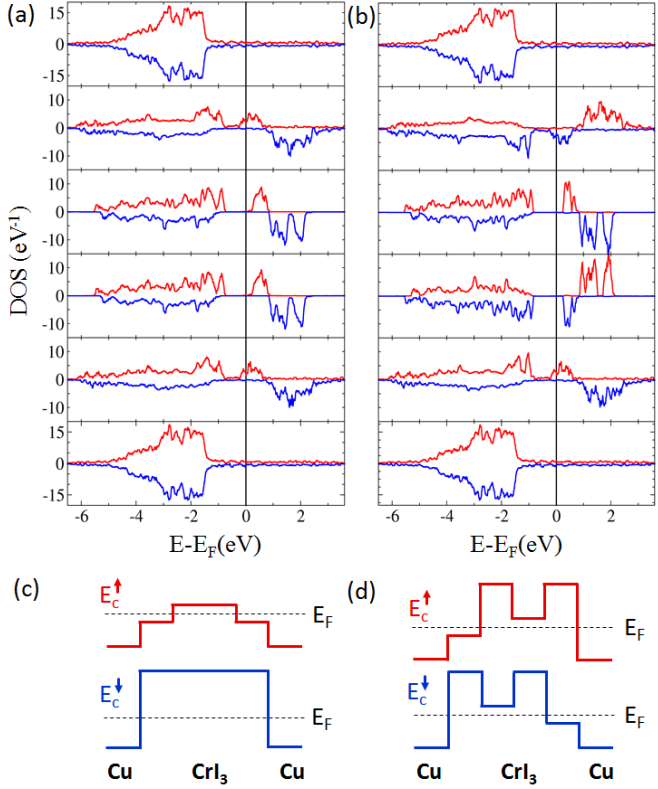

**Fig. 3** (a, b) Calculated layer- and spin-resolved density of states (DOS) in a Cu(5ML)/CrI$_3$(4ML)/Cu(5ML) tunnel junction as a function of energy for FM- (a) and AFM- (b) ordered CrI$_3$. The top and bottom panels display DOS of the interfacial Cu MLs, central panels display DOS of the four CrI$_3$ MLs. Majority (minority)-spin DOS are indicated by red (blue) lines. (c, d) Schematics of the spin-dependent potential barrier profile across the junction for FM- (c) and AFM- (d) ordered CrI$_3$. Majority (minority)-spin channels are indicated by red (blue) lines.

Calculations are performed as described in Methods. The atomic relaxation involves two steps. First, we find the minimum-energy distance, $d_{min}$, between the Cu (111) slab strained to the CrI$_3$ lattice constants and the CrI$_3$ (0001) slab, while keeping all the internal coordinates fixed, corresponding to bulk CrI$_3$ and Cu. Then, we fix $d_{min}$ and relax all the atomic coordinates in the Cu/CrI$_3$/Cu supercell. We find that at the interface Cu and I atoms form bonds with the bond length of 2.67 Å. This is not surprising, as iodine is known to form compounds with the noble metals, such as Pt, Au, and Ag. In addition, the relaxation slightly reduces the interlayer spacing between the CrI$_3$ monolayers to 6.43 Å, as compared to the bulk value of 6.45 Å. It also makes the Cr-I bonds asymmetric with a longer bond of 2.77 Å next to the interface and a shorter of 2.69 Å bond away from the interface, as compared to the bulk Cr-I bond length of 2.71 Å. These bonds are largely symmetric in the middle CrI$_3$ monolayers. A tunnel junction with a two-monolayer CrI$_3$ barrier shows similar relaxation behavior.[24]

**Cu/CrI$_3$/Cu tunnel junction: Electronic structure**
Electronic and transport properties of the Cu/CrI$_3$/Cu tunnel junction depend critically on the band offset between CrI$_3$ and Cu. Hence, we first calculate the band offset by analyzing the relative positions of the Fermi level in the Cu (111) slab and the VBM in the CrI$_3$ (001) slab when the vacuum potentials of the two slabs are matched. Figure 2c shows that the Fermi level in Cu (111) lies about 0.2 eV below the CBM of CrI$_3$, indicating that CrI$_3$ layer is expected to produce a tunneling barrier when is used in a Cu/CrI$_3$/Cu tunnel junction.

This fact agrees with the calculated layer-resolved density of states (DOS) for a Cu/CrI$_3$/Cu tunnel junction with four monolayers (MLs) of CrI$_3$ (Fig. 2a). As seen from Figures 3a and 3b, away from the interface, the Fermi level lies within the band gap of CrI$_3$ both for FM- (Fig. 3a) and AFM-ordered (Fig. 3b) CrI$_3$. Therefore, a tunnel barrier exists both for majority- and minority-spin electrons. However, the height of the barrier and its profile across the Cu/CrI$_3$/Cu tunnel junction are spin-dependent and different for FM- and AFM-ordered CrI$_3$. In the FM configuration, there is a uniform barrier across the junction with the spin-dependent barrier height of about 0.2 eV in the majority-spin channel and about 0.8 eV in the minority-spin channel (Fig. 3c). In this case, the transmission is expected to be large and dominated by majority-spin electrons. In the AFM configuration, the barrier height changes between low and high values on each consecutive CrI$_3$ monolayer for both spin channels (Fig. 3d). In this case, the conductance is expected to be low in each spin channel and largely spin independent due to the mirror-symmetric barrier profiles between the two spin channels.

There is another important feature in the electronic structure of the Cu/CrI$_3$/Cu tunnel junction. There is a sizable charge transfer from Cu to the first monolayer of CrI$_3$ at the Cu/CrI$_3$ interface. This is evident from the Fermi energy located within the majority-spin $e_g$ conduction band of the interfacial CrI$_3$ monolayer (Fig. 2a). Such an electron doping effect makes this CrI$_3$ monolayer "half-metallic," so that the majority-spin carriers experience no barrier within this monolayer (Fig. 3c, top panel). This feature of the electronic band structure at the Cu/CrI$_3$ interface adds additional contribution to the spin-filtering effect (and, in fact, dominates in the 2ML CrI$_3$ tunnel junction[24]), as discussed below.

**Cu/CrI$_3$/Cu tunnel junction: Transmission**
The transmission function is calculated as described in Methods by considering the Cu/CrI$_3$/Cu (111) supercell (Fig. 2a) as the scattering region ideally attached on both sides to semi-infinite Cu (111) electrodes. Figure 4a shows the calculated spin-resolved transmission $T_\sigma(E)$ ($\sigma = \uparrow, \downarrow$) across the FM-ordered Cu/CrI$_3$/Cu tunnel junction as a function of energy. We find that there is several orders of magnitude difference in the transmission values between the majority- and minority-spin conduction channels over the whole range of energies. Relevant to experiment, the transmission value at the Fermi energy $T_\sigma(E_F)$ is four orders of magnitude higher for the majority-spin electrons as compared to the minority-spin electrons. The spin polarization of the tunneling conductance SP = $(T_\uparrow - T_\downarrow)/(T_\uparrow + T_\downarrow)$ is virtually 100%. This manifests a perfect spin-filtering effect resulting from the spin-dependent potential barrier in CrI$_3$ (Fig. 3c). A qualitatively similar effect is predicted for the 2ML CrI$_3$ tunnel junction (Fig. S1c).

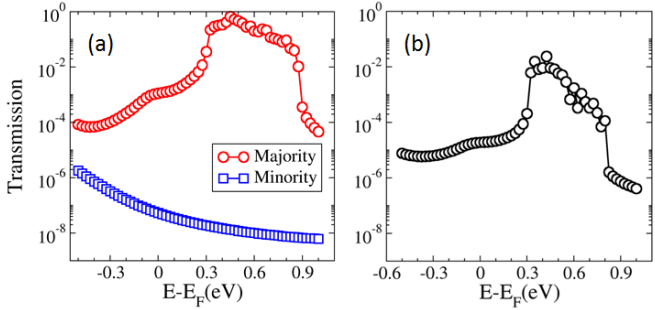

**Fig. 4** Spin-resolved transmission per lateral unit cell area across the Cu/CrI$_3$/Cu tunnel junction as a function of energy for FM- (a) and AFM- (b) ordered CrI$_3$.

Spin filtering is also responsible for the large TMR effect associated with the change in the magnetic ordering of CrI$_3$ by an applied magnetic field. For FM-aligned CrI$_3$, the total transmission, $T_{FM} = T_\uparrow + T_\downarrow$, is nearly the same as the majority-spin transmission $T_\uparrow$ (due to $T_\uparrow \gg T_\downarrow$), which is shown in Fig. 4a by red dots. For AFM-aligned CrI$_3$, the total transmission, $T_{AFM}$, is twice that per spin (due to symmetry of the considered tunnel junction making the spin transmission for each spin channel to be equal), which is shown in Fig. 4b. By comparing Figs. 4a and 4b, it is seen that $T_{AFM}$ as a function of energy exhibits similar features as $T_{FM}$, but $T_{AFM}$ is more than an order of magnitude lower than $T_{FM}$ over the whole range of energies. This is explained by the spin-dependent potential profiles shown in Figs. 3c and 3d for the FM- and AFM-ordered tunnel junction, respectively. While $T_{FM}$ is dominated by the majority-spin electrons encountering a low potential barrier across the whole CrI$_3$ layer (Fig. 3c, top panel), the AFM ordering makes the tunneling barrier high for either spin channel in those regions where the tunneling electrons have their spin antiparallel to the magnetic moment of the CrI$_3$ monolayer (Fig. 3d). Thus, while the spin filtering selects a highly transmittive majority-spin channel for the FM configuration, it suppresses the transmission in either spin channel for the AFM configuration by the consecutive filtering of the majority- and minority-spin carriers when they are tunneling across the barrier. The TMR value calculated at the Fermi energy is ($T_{FM} - T_{AFM})/T_{AFM}$ ~ 3,000%, which reveals a very large magnetoresistive effect in line with the experimental observations.[14-18] For the 2ML CrI$_3$ tunnel junction, we find that the TMR value is reduced to about 1,500%.[24] With increasing the CrI$_3$ barrier width, the TMR value is expected to grow exponentially.

As seen from Figures 3a and 3b, the spin filtering and TMR effects become even larger at higher energies. At about 0.3 eV above the Fermi energy, the majority-spin transmission across the FM-ordered CrI$_3$ is strongly enhanced (Fig. 3a). A similar feature is seen for either-spin transmission across the AFM-ordered CrI$_3$. This behavior is due to the electron energy entering the majority-spin conduction band of CrI$_3$. At these energies, the enhanced spin filtering and TMR result from the essentially "metallic" conduction of the majority-spin electrons, in contrast to the tunneling conduction of the minority-spin electrons. The predicted enhancement of the spin filtering at higher energies is likely responsible for the experimentally observed enhancement of TMR with applied bias voltage.[3,16,18] Similar behavior is predicted for a tunnel junction with a two-monolayer CrI$_3$ barrier (Fig. S1).

To obtain a deeper insight into the mechanism of spin-dependent tunneling, we calculate transmission $T(\mathbf{k}_\parallel)$ for different wave vectors $\mathbf{k}_\parallel$ in the two-dimensional Brillouin zone. Due to periodicity of the system in the plane perpendicular to the transport direction, wave vector $\mathbf{k}_\parallel$ in that plane is conserved during the tunneling process. Figures 4 a-c show the calculated spin- and $\mathbf{k}_\parallel$-resolved transmission across the Cu/CrI$_3$/Cu tunnel junction for FM (Figs. 4 a,b) and AFM (Fig. 4c) aligned CrI$_3$. As is seen from the figures, in all cases the transmission is

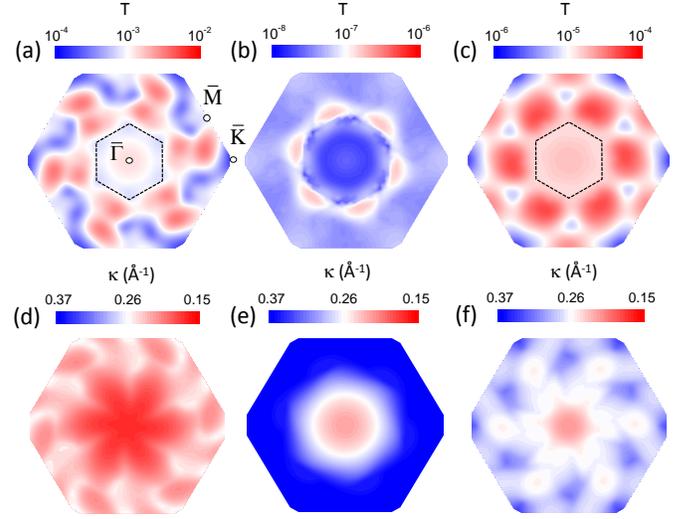

**Fig. 5** (a-c) $\mathbf{k}_\parallel$-resolved transmission ($T$) across the Cu/CrI$_3$/Cu tunnel junction in the 2D Brillouin zone for majority (a) and minority (b) spin for FM and AFM (c) ordered CrI$_3$. The transmission is plotted in a logarithmic scale (note different scales for (a), (b), and (c)). The high symmetry points in the 2D Brillouin zone are shown. Dashed lines in (a) and (c) indicate the region where there are no available propagating states in the interface monolayer of CrI$_3$. (d-f) $\mathbf{k}_\parallel$-resolved lowest decay rate for majority (a) and minority (b) spin for FM and AFM (c) ordered bulk CrI$_3$ calculated at 0.2eV below the CBM.

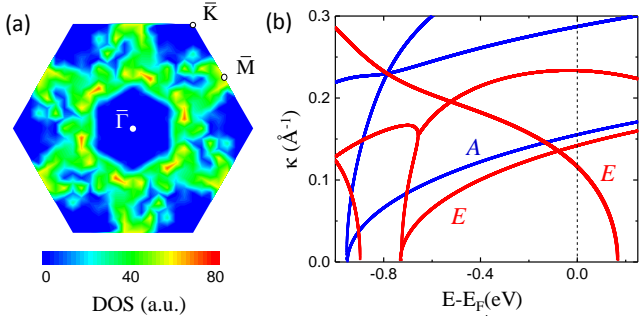

**Fig. 6** $k_\parallel$-resolved majority-spin density of states (DOS) of bulk FM-ordered CrI$_3$ plotted in the 2D Brillouin zone at $E = 0.23$ eV above the CBM (a). The complex band structure of CrI$_3$ as a function of energy for majority- (red curves) and minority- (blue curves) spin electrons. The dashed line indicates the Fermi energy in the Cu/CrI$_3$/Cu tunnel junction.

somewhat reduced around the high symmetry $\bar{\Gamma}$ point and enhanced in certain peripheral regions of the 2D Brillouin zone.

It order to understand this behavior, we analyze the complex band structure of bulk CrI$_3$. It is known that tunneling through insulators can be interpreted in terms of evanescent states,[25] and the method to investigate them is the complex band structure in the energy gap region.[26-28] To obtain the complex band structure, for each $\mathbf{k}_\parallel$, we calculate the dispersion relation $E = E(k_z)$, allowing complex $k_z = q + i\kappa$. The imaginary part $\kappa$ is the decay rate, so that the corresponding wave functions decay as $\sim e^{-\kappa z}$. The evanescent states with the lowest decay rate are expected to be most efficient in the tunneling process and hence to control the transmission. In Figs. 5d-f, we plot the lowest decay rate $\kappa(\mathbf{k}_\parallel)$ for the energy in the band gap of bulk CrI$_3$, corresponding to the Fermi energy in the Cu/CrI$_3$/Cu tunnel junction. Below we discuss the relationship between the $\mathbf{k}_\parallel$ dependent transmission (Figs. 5a-c) and the lowest decay rate (Figs. 5d-f).

First, we consider the majority-spin conduction channel for the FM aligned CrI$_3$. Comparing Figures 5a and 5d, we see that away from $\bar{\Gamma}$ point closer to the edges of the 2D Brillouin zone (beyond the hexagon indicated by the dashed line in Fig. 5a), there is a clear resemblance between the high values of $T(\mathbf{k}_\parallel)$ and the low decay rates $\kappa(\mathbf{k}_\parallel)$. This behavior is consistent with the expectation that the slowest decaying evanescent states produce the highest transmission. However, the six-petal flower feature with the lowest decay rate centered at the $\bar{\Gamma}$ point seen in the color plot of $\kappa(\mathbf{k}_\parallel)$ (Fig. 5d) is not reproduced in the color plot of $T(\mathbf{k}_\parallel)$ (Fig. 5a).

This behavior can be understood in terms of the $\mathbf{k}_\parallel$ dependent density of states (DOS) at the interfacial monolayer of CrI$_3$. As was discussed in conjunction with the DOS plotted in Figure 3a, due to the charge transfer from Cu, this monolayer of CrI$_3$ has a non-zero majority-spin DOS at the Fermi energy and thus serves as the termination of the metal electrode rather than a tunneling barrier. The $\mathbf{k}_\parallel$ dependent DOS determines the efficiency of transmission across this monolayer from the Cu electrode further into the tunneling region of CrI$_3$. Figure 5a shows the $\mathbf{k}_\parallel$ dependent majority-spin DOS of bulk CrI$_3$ at $E = 0.23$ eV above the CBM corresponding to the Fermi energy in the interfacial CrI$_3$ monolayer (see Fig. 3a). The key feature of this plot is the absence of electronic states within the hexagon region around the $\bar{\Gamma}$ point. Thus, while the electronic states are available in the most peripheral region of the 2D Brillouin zone, supporting the metallic-like transmission across this area, there are no propagating state in the central hexagon region, which implies that the transmission though this area is suppressed. This explains the reduced transmission seen within the hexagon region in Figure 5a indicated by the dashed line. The similar behavior is observed for the Cu/CrI$_3$(2ML)/Cu tunnel junction (Fig. S1d). The six-petal flower feature, showing the lowest decay rates in CrI$_3$ (Fig. 5d), reveals itself only at the edges of the petals where the enhanced transmission is seen in Fig. 5a.

For AFM-ordered CrI$_3$, the $\mathbf{k}_\parallel$-resolved transmission across the Cu/CrI$_3$/Cu tunnel junction (Fig. 5c) shows a qualitatively similar behavior, although it is lower by about two orders in magnitude than the transmission for FM-ordered CrI$_3$. Comparing Figures 5c and 5f, we see a qualitative similarity between the high values of $T(\mathbf{k}_\parallel)$ and the low decay rates $\kappa(\mathbf{k}_\parallel)$ in the peripheral region of the 2D Brillouin zone. The transmission if suppressed however within the hexagon area (indicated by the dashed line in Fig. 5c) due to the absence of electronic states at one the interfaces (left or right depending on the spin channel) in this region of the Brillouin zone (Fig. 6a).

The situation is different for the minority-spin transmission $T(\mathbf{k}_\parallel)$ across the FM-ordered Cu/CrI$_3$/Cu tunnel junction. First, as seen from Figure 5b, the transmission values are about four orders in magnitude smaller than those for the majority spin (Fig. 5a). Second, there is virtually no correlation between $T(\mathbf{k}_\parallel)$ and $\kappa(\mathbf{k}_\parallel)$. According to Figure 5e, the expected high transmission should occur at the $\bar{\Gamma}$ point. However, as seen from Figure 5b, transmission is suppressed in the hexagon area around this point. This disparity is due to the symmetry mismatch between the states available at the Fermi energy in Cu and the evanescent states in CrI$_3$ at the $\bar{\Gamma}$ point. This can be understood from Figure 6b, where we plot the imaginary part of the wave vector $\kappa$ in bulk CrI$_3$ as a function of energy at the $\bar{\Gamma}$ point. There are several majority- and minority-spin evanescent bands crossing the Fermi energy (indicated by the vertical dashed line in Fig. 6b). The states with the lowest decay rate belong to the *E* symmetry representation for majority spin and the *A* symmetry representation for minority spin, corresponding the group of the wave vector along the Γ-Z direction in the Brillouin zone of bulk CrI$_3$. In terms of atomic orbitals, the *A* symmetry state transforms as *s*, $p_z$ and $d_{z2}$ orbitals, while the *E* symmetry state transforms as $p_x$, $p_y$, $d_{x2-y2}$, $d_{xy}$, $d_{yz}$, and $d_{xz}$ orbitals. The incoming Cu states at the $\bar{\Gamma}$ point contain mixture of the Cu *s, p,* and *d* atomic orbitals; however, the contribution of the *A* symmetry states is much smaller (~5%) compared to the contribution of the *E* symmetry states. Thus, the symmetry mismatch between the in-coming and evanescent states suppresses the transmission around the $\bar{\Gamma}$ point and makes the overall transmission of the minority-spin electrons low.

**Summary**

Using first-principles methods based on density functional theory, we have explored the mechanisms of the spin filtering effect in tunnel junctions with a 2D ferromagnetic insulator CrI$_3$ used as a tunneling barrier. Our results demonstrate that a four-monolayer CrI$_3$ barrier provides virtually a fully spin-polarized transmission in a tunnel junction utilizing fcc Cu (111) electrodes. The predicted TMR effect is about 3,000%, associated with the change of magnetic ordering in CrI$_3$ from AFM to FM. With increasing the CrI$_3$ barrier width, the TMR ratio is expected to grow exponentially. A simple qualitative picture of the spin filtering effect is based on the exchange splitting of the spin bands in CrI$_3$ producing different barrier heights for majority- and minority-spin carriers. The detailed analysis of the spin-dependent electronic band structure and the spin and wave-vector resolved transmission in Cu/CrI$_3$/Cu tunnel junctions show however a more multifaceted behavior. We find, in particular, that there is a sizable charge transfer from Cu to the interfacial monolayer of CrI$_3$ at the Cu/CrI$_3$ interface. This makes this CrI$_3$ monolayer "half-metallic," i.e. such that the majority spins experience no barrier within this monolayer, whereas the minority spins encounter a tunnel barrier. Furthermore, the detailed analysis of the complex band structure of CrI$_3$ shows that the spin filtering effect cannot be quantitatively understood entirely based on a simple model of the spin-dependent barrier heights. The important roles are played by the $\mathbf{k}_\parallel$- and spin-dependent evanescent states, their symmetries and matching to the propagating states in the electrodes. These effects add new important features to the mechanism of spin-filtering in CrI$_3$ tunnel junctions. Our results provide important insights into the understanding of spin-filtering by magnetic insulators, such as CrI$_3$, and may be useful for the design of novel magnetoresistive devices based on 2D magnetic crystals.

**Methods**

The atomic and electronic structure of the system is calculated using density functional theory (DFT) as implemented in Vienna *ab initio* simulation package (VASP).[29,30] The projected augmented plane wave (PAW) method is used to approximate the electron-ion potential.[31] The exchange-correlation potential is described within generalized density approximation (GGA). In the calculations, we use a kinetic energy cutoff of 340 eV for the plane wave expansion of the PAWs and 6×6×1 and grid of *k* points for the Brillouin zone integration. For the accurate description of the interlayer separation in CrI$_3$, the van der Waal's interaction is included in the calculations.[32] The in-plane lattice constant is fixed to the theoretical value of the bulk CrI$_3$ lattice constant of 6.81 Å, which is close to the experimental value of 6.88 Å. With this in-plane lattice constant, the out-of-plane lattice constant and all the internal coordinates of the supercells are relaxed until the Hellman-Feynman force on each atom is less than 0.01 eV/Å. Under these conditions, the inter-plane separation of the CrI$_3$ monolayers is found to be 6.45 Å. All the calculations are spin-polarized with the Cr moment aligned parallel within the CrI$_3$ monolayers. The FM (AFM) phase of CrI$_3$ is described by assuming parallel (antiparallel) alignment of the magnetic moments of the adjacent CrI$_3$ monolayers. For the AFM structure, we increase the lattice parameter by a factor of two in the *c*-direction to make the CrI$_3$ structure compatible with the periodic boundary conditions. The band offset calculations are performed using 6 layers thick Cu slab (equivalent to two unit cells of Cu(111)) padded with 16 Å vacuum and 8 monolayer-thick AFM-ordered CrI$_3$ slab padded with 16 Å of vacuum. The work function is defined by the potential energy difference between the vacuum and the Fermi level for the Cu slab and the valence band maximum for the CrI$_3$ slab.

The transmission and complex band structure calculations are carried out using the Atomistic Simulation Toolkit (ATK) with the PBE pseudopotentials distributed in the QuantumWise package (Version 2015.1).[33,34] As implemented in the code, the transmission is calculated using the non-equilibrium Green's function approach.[35] The Cu/CrI$_3$/Cu (111) supercell is used as the scattering region, ideally attached on both sides to semi-infinite Cu (111) leads.[33] This structure has open boundary conditions in the *c*-direction but is periodic in the *a-b* plane. The latter property makes the in-plane Bloch wave vector $\mathbf{k}_\parallel$ a good quantum number, so that transmission *T* is a function of $\mathbf{k}_\parallel$. For calculating transmission, the two-dimensional Brillouin zone is sampled using a uniform 51×51 $\mathbf{k}_\parallel$ mesh.

The atomic structure of the Cu/CrI$_3$/Cu(111) junctions is designed with the help VNL distributed with Atomistic toolkit.[36]

**Acknowledgments**

The authors thank Dr. Lingling Tao for valuable discussions. This work was supported by the National Science Foundation (NSF) through Nebraska Materials Research Science and Engineering Center (MRSEC) (NSF Grant No. DMR-1420645). Computations were performed at the University of Nebraska Holland Computing Center. The atomic structure was produced using VESTA software[37].

1. Geim, A. K.; Grigorieva, I. V. Van der Waals heterostructures. *Nature* **2013**, *499* (7459), 419–425.
2. Novoselov, K. S.; Mishchenko, A.; Carvalho, A.; Castro Neto, A. H. 2D Materials and van der Waals heterostructures. *Science* **2016**, *353* (6298), aac9439.
3. Huang, B.; Clark, G.; Navarro-Moratalla, E.; Klein, D. R.; Cheng, R.; Seyler, K. L.; Zhong, D.; Schmidgall, E.; McGuire, M. A.; Cobden, D. H. Layer-dependent ferromagnetism in a van der Waals crystal down to the monolayer limit. *Nature* **2017**, *546* (7657), 270–273
4. Gong, C.; Li, L.; Li, Z.; Ji, H.; Stern, A.; Xia, Y.; Cao, T.; Bao, W.; Wang, C.; Wang, Y.; Qui. Z. Q; Cava, R. J.; Louie, S. G.; Xia, J.; Zhang, X. Discovery of intrinsic ferromagnetism in two-dimensional van der Waals crystals. *Nature* **2017**, *546* (7657), 265–269.
5. Bonilla, M.; Kolekar, S.; Ma, Y.; Diaz, H. C.; Kalappattil, V.; Das, R.; Eggers, T.; Gutierrez, H. R.; Phan, M.-H.; Batzill, M. Strong room-temperature ferromagnetism in VSe$_2$ monolayers on van der Waals substrates. *Nat. Nanotechnol.* **2018**, *13* (4), 289–293.
6. Mermin, N. D.; Wagner, H. Absence of ferromagnetism or antiferromagnetism in one- or two-dimensional isotropic Heisenberg models. *Phys. Rev. Lett.* **1966**, 17 (26), 1307–1307.
7. Zhang, W.-B.; Qu, Q.; Zhu, P.; Lam, C.-H. Robust intrinsic ferromagnetism and half semiconductivity in stable two-dimensional single-layer chromium trihalides. *J. Mater. Chem. C* **2015**, 3 (48), 12457–12468.
8. McGuire, M. A.; Dixit, H.; Cooper, V. R.; Sales, B. C. coupling of crystal structure and magnetism in the layered, ferromagnetic insulator CrI$_3$. *Chem. Mater.* **2015**, *27* (2), 612–620.
9. Dillon, J. F.; Olson, C. E. Magnetization, resonance, and optical properties of the ferromagnet CrI$_3$. *J. Appl. Phys.* **1965**, *36* (3), 1259–1260.
10. Lado, J. L.; Fernández-Rossier, J. On the origin of magnetic anisotropy in two dimensional CrI$_3$. *2D Mater.* **2017**, *4* (3), 035002.
11. Jiang, S.; Li, L.; Wang, Z.; Mak, K. F.; Shan, J. Controlling magnetism in 2D CrI$_3$ by electrostatic doping. *Nat. Nanotechnol.* **2018**, *13* (7), 549–553.
12. Jiang, S.; Shan, J.; Mak, K. F. Electric-field switching of two-dimensional van der Waals magnets. *Nat. Mater.* **2018**, *17* (5), 406–410.
13. Huang, B.; Clark, G.; Klein, D. R.; Macneill, D.; Navarro-Moratalla, E.; Seyler, K. L.; Wilson, N.; McGuire, M. A.; Cobden, D. H.; Xiao, D.; Yao, W.; Jarillo-Herrero, P. Electrical control of 2D magnetism in bilayer CrI$_3$. *Nat. Nanotechnol.* **2018**, *13* (7), 544–548.
14. Klein, D. R.; MacNeill, D.; Lado, J. L.; Soriano, D.; Navarro-Moratalla, E.; Watanabe, K.; Taniguchi, T.; Manni, S.; Canfield, P.; Fernández-Rossier, J.; Jarillo-Herrero, P. Probing magnetism in 2D van der Waals crystalline insulators via electron tunneling. *Science.* **2018**, *360* (6394), 1218–1222.
15. Tsymbal, E. Y.; Mryasov, O. N.; LeClair, P. R. Spin-dependent tunnelling in magnetic tunnel junctions. *J. Phys. Condens. Matter* **2003**, *15* (4), R109–R142.
16. Song, T.; Cai, X.; Tu, M. W.-Y.; Zhang, X.; Huang, B.; Wilson, N. P.; Seyler, K. L.; Zhu, L.; Taniguchi, T.; Watanabe, K.; Mcguire, M.A.; Cobden, D. H.; Xiao, D.; Yao, W.; Xu, X. Giant tunneling magnetoresistance in spin-filter van der Waals heterostructures. *Science* **2018**, *360* (6394), 1214–1218.
17. Wang, Z.; Gutiérrez-Lezama, I.; Ubrig, N.; Kroner, M.; Gibertini, M.; Taniguchi, T.; Watanabe, K.; Imamoğlu, A.; Giannini, E.; Morpurgo, A. F. Very large tunneling magnetoresistance in layered magnetic semiconductor CrI$_3$. *Nat. Commun.* **2018**, 9 (1), 2516.
18. Kim, H. H.; Yang, B.; Patel, T.; Sfigakis, F.; Li, C.; Tian, S.; Lei, H.; Tsen, A. W. One million percent tunnel magnetoresistance in a magnetic van der Waals heterostructure. *Nano Lett.* **2018**, *18* (8), 4885–4890.
19. Wang, H.; Eyert, V.; Schwingenschlögl, U. Electronic structure and magnetic ordering of the semiconducting chromium trihalides CrCl$_3$, CrBr$_3$, and CrI$_3$. *J. Phys. Condens. Matter* **2011**, *23* (11), 116003.
20. Santos, T.; Moodera, J. Spin-filter tunneling. In *Handbook of Spin Transport and Magnetism*; edited by Tsymbal, E. Y. and Žutić, I. **2012**, Chapter *13*, pp. 251–266.
21. Lukashev, P. V.; Burton, J. D.; Smogunov, A.; Velev, J. P.; Tsymbal E. Y. Interface states in CoFe$_2$O$_4$ spin-filter tunnel junctions. *Phys. Rev. B* **2017**, *88* (13), 134430.
22. Approximating the exchange-correlation potential with a semi-local functional, such as HSE, overestimates the band gap up to about 1.6 eV. See, e.g., Heyd, J.; Scuseria, G. E.; Ernzerhof, M. Hybrid functionals based on a screened Coulomb potential. *J. Chem. Phys.* **2003**, *118* (18), 8207–8215. *J. Chem. Phys.* **2006**, *124* (21), 219906 and Krukau, A. V.; Vydrov, O. A.; Izmaylov, A. F.; Scuseria, G. E. Influence of the exchange screening parameter on the performance of screened hybrid functionals. *J. Chem. Phys.* **2006**, *125* (22), 224106.
23. Jelver, L.; Larsen, P. M.; Stradi, D.; Stokbro, K.; Jacobsen, K. W. Determination of low-strain interfaces via geometric matching. *Phys. Rev. B* **2017**, *96* (8), 085306.
24. Supplementary Information.
25. Heine, V. On the general theory of surface states and scattering of electrons in solids. *Proc. Phys. Soc.* **1963**, *81* (2), 300–310; *Ibid* Theory of surface states. *Phys. Rev.* **1965**, *138* (6A), A1689–A1696.
26. Mavropoulos, P.; Papanikolaou, N.; Dederichs, P. H. Complex band structure and tunneling through ferromagnet/insulator/ferromagnet junctions. *Phys. Rev. Lett.* **2000**, *85* (5), 1088–1091.
27. Velev, J. P.; Belashchenko, K. D.; Jaswal, S. S.; Tsymbal, E. Y. Effect of oxygen vacancies on spin-dependent tunneling in Fe∕MgO∕Fe magnetic tunnel junctions. *Appl. Phys. Lett.* **2007**, *90* (7), 072502.
28. Lukashev, P.; Wysocki, A.; Velev, J. P.; van Schilfgaarde, M.; Jaswal, S. S.; Belashchenko, K. D.; Tsymbal E. Y. Spin filtering with EuO: Insight from a complex band structure. *Phys. Rev. B* **2012**, *85* (22), 224414.
29. Kresse, G.; Joubert, D. From ultrasoft pseudopotentials to the projector augmented-wave method. *Phys. Rev. B* **1999**, *59*, 1758–1775.
30. Kresse, G.; Furthmüller, J. Efficient iterative schemes for ab initio total-energy calculations using a plane-wave basis set. *Phys. Rev. B* **1996**, *54*, 11169–11186.
31. Blöchl, P. E. Projector augmented-wave method. *Phys. Rev. B* **1994**, *50*, 17953–17979.
32. Grimme, S.; Antony, J.; Ehrlich, S.; Krieg, H. A Consistent and accurate *ab initio* parametrization of density functional dispersion correction (DFT-D) for the 94 elements H–Pu. *J. Chem. Phys.* **2010**, *132* (15), 154104; Grimme, S.; Ehrlich, S.; Goerigk, L. Effect of the damping function in dispersion corrected density functional theory. *J. Comput. Chem.* **2011**, *32* (7), 1456–1465.

[33] Brandbyge, M.; Mozos, J.-L.; Ordejón, P.; Taylor, J.; Stokbro, K. Density-functional method for nonequilibrium electron transport. *Phys. Rev. B* **2002**, *65* (16), 165401.

[34] Soler, J. M.; Artacho, E.; Gale, J. D.; García, A.; Junquera, J.; Ordejón, P.; Sánchez-Portal, D. The SIESTA method for *ab initio* order-*N* materials simulation. *J. Phys. Condens. Matter* **2002**, *14* (11), 2745–2779.

[35] Taylor, J.; Guo, H.; Wang, J. *Ab Initio* modeling of quantum transport properties of molecular electronic devices. *Phys. Rev. B* **2001**, *63* (24), 245407.

[36] Atomistix Toolkit version 2015.1., Synopsys QuantumWise A/S ([www.quantumwise.com](www.quantumwise.com)).

[37] Momma, K.; Izumi, F.; VESTA3 for three-dimensional visualization of crystal, volumetric and morphology data. *J. Appl. Crystallogr.* **2011**, *44* (6), 1272–1276.